# Power Factor Angle Droop Control—A General Decentralized Control of Cascaded inverters

Yao Sun, Lang Li, Guangze Shi, Xiaochao Hou, Mei Su

[1]***Abstract*—This letter proposes a general decentralized control of cascaded inverters-power factor angle droop control. Compared to the existing control strategies, it has the following attractive benefits: 1) it is suitable for both grid-connected and islanded modes; 2) Seamless transition between different modes can be obtained; 3) stability condition in the grid-connected mode is independent of the transmission line impedance; 4) it is suited for any types of loads in islanded modes; 5) multi-equilibrium point problem is avoided; 6) it is suitable for four quadrant operation. The small signal stability of the control is proved. And the feasibility of the proposed method is verified by simulation.**

## I. Introduction

Parallel and series (cascade) operations are two important ways to form large-scale power systems. Generally speaking, the parallel operation is more widely used due to its ease of use and reliability. For example, lots of micro-grid are formed by a large number of inverters in parallel. However, series operation is indispensable to form a high-voltage equipment or network. For instance, the cascaded inverters are used for high-voltage motor divers, STATCOM, and energy storage system [1-2].

In the past, most of the control schemes of cascaded inverters pertain to the centralized control [3-4]. However, the centralized control depends on real-time communication networks and powerful centralized controller, which often lead to the reduced reliability due to communication failure, and higher capital costs. Moreover, its implementation will become more difficult when the number of the inverter modules is large and the distance between the modules is longer.

Recently decentralized strategies of cascaded inverters have received wide attention because they could overcome the drawbacks of the centralized control methods mentioned above [5-9]. For the cascaded inverters in islanded mode, [5] firstly presents an inversed power factor droop control, which is suited for only resistance inductance (RL) loads. To broaden application scope of the method [5], f-P/Q method is proposed [6]. This control is suitable for both RL and resistance capacitance (RC) loads. However, it is still unfeasible for pure resistance load at least from the mathematical point of view. Meanwhile, it has the problem of multiple equilibrium points, which may bring about some undesired operating states. Further, [7] proposes an improved decentralized control with unique equilibrium point. Contrast to the islanded mode, the situation in the grid-connected mode is different. Literature [8] proposes a fully decentralized control, which is the first attempt to control the cascaded inverter in the decentralized manner. Further the same author proposes a reactive power-frequency droop control scheme for the application in reactive power compensation [9]. However, they are only suited for some specific transmission line types. Literature [10] presents a distributed power control for grid-tied photovoltaic generation. However, the PCC voltage information must be available for each modules, which increases the implementation difficulty of the method. In addition to the decentralized control schemes only for islanded or grid-connected mode, a decentralized control suited for both of them is desired in practice.

To address the concerns above, this letter proposes a general decentralized control of cascaded inverters for both grid-connected and islanded modes operation. Compared to the existing methods, the proposed power factor angle droop control has the following features:

- *A unified control scheme*. The methods proposed in [5-10] are only applied to the grid-connected mode or islanded mode. Alternatively, the proposed scheme is a unified control scheme for both the two modes. Thus, seamless transition between them can be obtained.
- *Suited for all types of loads*. The method in [5] is only suited for RL loads, and the methods in [6-7] are feasible for RL and RC loads. However, the pure resistance load is not covered. Instead, the proposed scheme is a solution of all types of loads.
- *Unique equilibrium point*. The method in [6] has the problem of the multi-equilibrium points. The proposed one holds a unique equilibrium point.
- *Suited for all types of transmission line impedance*. The method in [8] is only applied to the inductive transmission line, while [9] for the resistive transmission line. However, the proposed one is feasible to all types of the transmission line impedance.
- *Four quadrant operation*. The methods in [5-10] could only work in parts of four quadrant operation. However, the proposed one can realize the four quadrant operation.

## II. Analysis of Proposed Control Strategy

### A. Configuration of cascaded inverter system

The structure of the cascaded inverters consisting of $n$ DG units is shown in Fig. 1. Different from the conventional cascaded inverter systems, the DG units may distribute broader areas, where the real-time communication is unavailable. The cascaded inverter system could operate in grid-connected or islanded mode by switching the static transfer switch (STS).

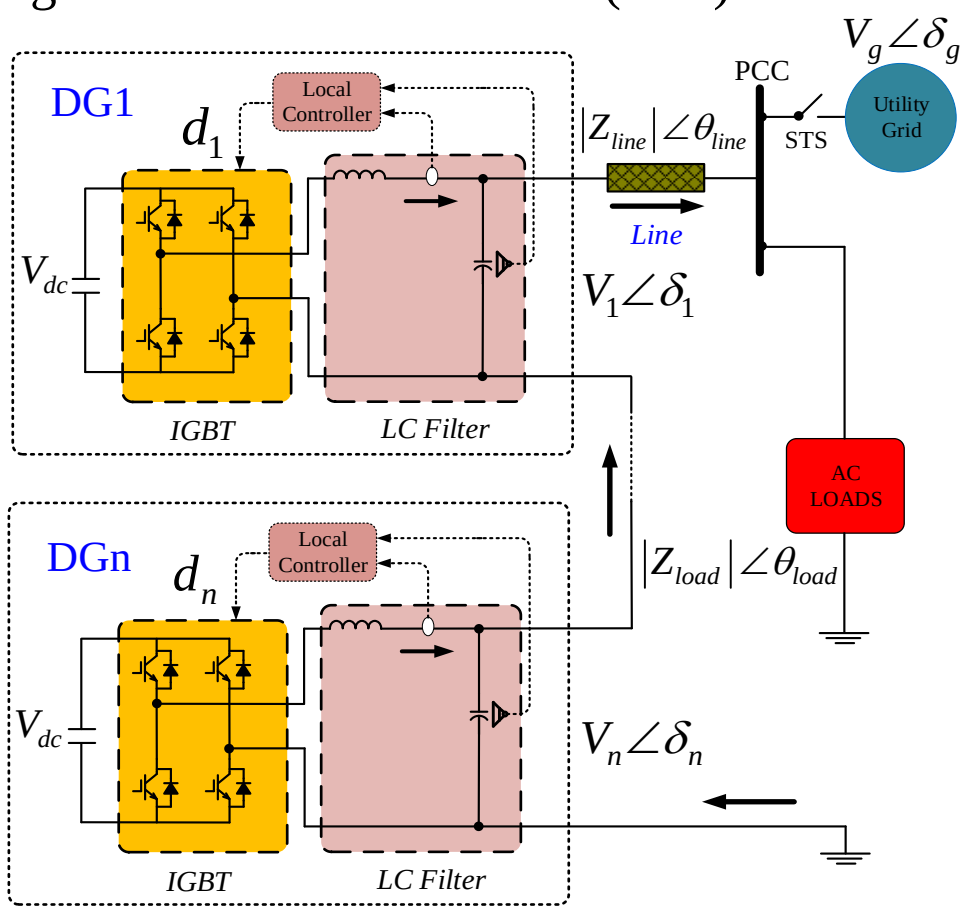

Fig. 1. Structure of the cascaded inverters.

### B. Power transmission characteristics

**In the islanded mode**, the output active power $P_i$ and reactive power $Q_i$ of the $i^{th}$ DG are derived as follows

$$P_i + jQ_i = V_i e^{j\delta_i}\left(\sum_{i=1}^{n} V_i e^{j\delta_i} \Big/ \left|Z'_{load}\right| e^{j\theta'_{load}}\right)^* \tag{1}$$

where $V_i$ and $\delta_i$ represent the voltage and phase angle of the $i^{th}$ unit. $Z'_{load}$ and $\theta'_{load}$ are the impedance and impedance angle of the generalized load, which includes transmission line and the load, respectively. Decomposing (1) into real and imaginary parts, the power transmission characteristic in the islanded mode is given by

$$P_i = \frac{V_i}{\left|Z'_{load}\right|}\sum_{j=1}^{n} V_j \cos\left(\delta_i - \delta_j + \theta'_{load}\right) \tag{2}$$

$$Q_i = \frac{V_i}{\left|Z'_{load}\right|}\sum_{j=1}^{n} V_j \sin\left(\delta_i - \delta_j + \theta'_{load}\right) \tag{3}$$

**In the grid-connected mode**, $P_i$ and $Q_i$ are expressed as

$$P_i + jQ_i = V_i e^{j\delta_i}\left(\left(\sum_{i=1}^{n} V_i e^{j\delta_i} - V_g e^{j\delta_g}\right) \Big/ \left|Z_{line}\right| e^{j\theta_{line}}\right)^* \tag{4}$$

where $Z_{line}\angle\theta_{line}$ represents the transmission line impedance. $V_g$ and $\delta_g$ represent the voltage amplitude and phase angle of the grid, respectively. The power transmission characteristic in the grid-connected mode is

$$P_i = \frac{V_i}{\left|Z_{line}\right|}\left(\sum_{j=1}^{n} V_j \cos\left(\delta_i - \delta_j + \theta_{line}\right) - V_g \cos\left(\delta_i - \delta_g + \theta_{line}\right)\right) \tag{5}$$

$$Q_i = \frac{V_i}{\left|Z_{line}\right|}\left(\sum_{j=1}^{n} V_j \sin\left(\delta_i - \delta_j + \theta_{line}\right) - V_g \sin\left(\delta_i - \delta_g + \theta_{line}\right)\right) \tag{6}$$

### C. Power Factor Angle Droop Control

The proposed power factor angle droop control strategy of the cascaded inverters is expressed as

$$\omega_i = \omega^* - m\left(\varphi_i - \varphi^*\right) \tag{7}$$

$$V_i = V^* \tag{8}$$

where $\omega_i$ is the angular frequency. $\omega^*$, $V^*$, $\varphi^*$ are the nominal angular frequency, voltage amplitude and power factor angle. $m$ is a positive coefficient, $\varphi_i$ is the power factor angle.

As seen, the proposed scheme in (7) and (8) only needs the local information of each module, thus it is a decentralized approach.

### D. Steady-state Analysis

In the steady-state, from (7), we have

$$\varphi_i = \varphi_j \tag{9}$$

where $i, j \in \{1, 2, \cdots, n\}$ .

In the islanded mode, according to (2), (3), (8) and (9), it is not difficult to obtain the following equations.

$$\begin{cases} P_i = P_j \\ Q_i = Q_j \end{cases} \tag{10}$$

In the grid-connected mode, the similar conclusions about the active and reactive power can be drawn as above.

### E. Small Signal Stability Analysis

To prove the stability of the proposed method in the islanded and grid-connected modes, the small signal analysis near the equilibrium point is carried out.

Assume $\delta_s$ is the synchronous phase-angle of cascaded inverters in the steady state, and denote $\tilde{\delta}_i = \delta_i - \delta_s$ . Since $\dot{\tilde{\delta}}_i = \omega_i$ , (7) is rewritten as

$$\dot{\tilde{\delta}}_i = \omega^* - m\left(\varphi_i - \varphi^*\right) \tag{11}$$

Linearizing (11) around the equilibrium point, we have

$$\Delta\dot{\tilde{\delta}}_i = -m\Delta\varphi_i \tag{12}$$

**In the islanded mode**, combining (2)and (3), yields

$$\varphi_i = \operatorname{atan}\frac{Q_i}{P_i} = \operatorname{atan}\frac{\sum_{j=1}^{n}\sin\left(\delta_i - \delta_j + \theta'_{load}\right)}{\sum_{j=1}^{n}\cos\left(\delta_i - \delta_j + \theta'_{load}\right)} \tag{13}$$

Then, linearizing (13), we have

$$\Delta\varphi_i = \frac{1}{n}\sum_{j=1}^{n}\left(\Delta\delta_i - \Delta\delta_j\right) \tag{14}$$

Combining (12) with (14) yields

$$\Delta\dot{\tilde{\delta}}_i = -\frac{m}{n}\sum_{j=1}^{n}\left(\Delta\delta_i - \Delta\delta_j\right) \tag{15}$$

Rewrite (15) as matrix form, we have

$$\dot{\boldsymbol{X}} = \boldsymbol{A}\boldsymbol{X} \tag{16}$$

where $\boldsymbol{X} = \left[\Delta\delta_1 \quad \cdots \quad \Delta\delta_n\right]^T$ , $\boldsymbol{A} = -\frac{m}{n}\boldsymbol{L}$ . $\boldsymbol{L}$ is a Laplacian matrix, and it is expressed as

$$\boldsymbol{L} = \begin{bmatrix} n-1 & -1 & \cdots & -1 \\ -1 & n-1 & \cdots & -1 \\ \vdots & \vdots & \ddots & \vdots \\ -1 & -1 & \cdots & n-1 \end{bmatrix} \tag{17}$$

The eigenvalues of $\boldsymbol{A}$ are expressed as

$$\lambda_1\left(\boldsymbol{A}\right) = 0, \lambda_2\left(\boldsymbol{A}\right) = \cdots = \lambda_n\left(\boldsymbol{A}\right) = -m \tag{18}$$

Clearly, the system in the islanded mode is stable [12]. Moreover, the stability does not depend on load parameters.

**In the grid-connected mode**, combining (5) with (6) yields

$$\varphi_i = \operatorname{atan}\frac{\sum_{j=1}^{n} V_j \sin\left(\delta_i - \delta_j + \theta_{line}\right) - V_g \sin\left(\delta_i - \delta_g + \theta_{line}\right)}{\sum_{j=1}^{n} V_j \cos\left(\delta_i - \delta_j + \theta_{line}\right) - V_g \cos\left(\delta_i - \delta_g + \theta_{line}\right)} \tag{19}$$

Linearization of (19) around the equilibrium point results in

$$\Delta\varphi = a\Delta\delta_i + b\sum_{j=1, i\neq j}^{n}\Delta\delta_j \tag{20}$$

where

$$a = \frac{\left(n^2 - n\right)\left(V^*\right)^2 + V_g^2 + (1-2n)V^* V_g \cos(\delta_s - \delta_g)}{n^2\left(V^*\right)^2 + V_g^2 - 2nV^* V_g \cos(\delta_s - \delta_g)} \tag{21}$$

$$b = \frac{V^* V_g \cos\left(\delta_s - \delta_g\right) - n\left(V^*\right)^2}{n^2\left(V^*\right)^2 + V_g^2 - 2nV^* V_g \cos(\delta_s - \delta_g)} \tag{22}$$

Substituting (20) into (12) yields

$$\dot{\boldsymbol{X}} = \boldsymbol{B}\boldsymbol{X} \tag{23}$$

where

$$\boldsymbol{B}=-m\begin{bmatrix} a & b & \cdots & b \\ b & a & \cdots & b \\ \vdots & \vdots & \ddots & \vdots \\ b & b & \cdots & a \end{bmatrix} \tag{24}$$

The eigenvalues of $\boldsymbol{B}$ are given by

$$\lambda_1(\boldsymbol{B})=-mV_g\left(V_g-nV^*\cos(\bar{\delta}-\delta_g)\right),\lambda_2(\boldsymbol{B})=\lambda_n(\boldsymbol{B})=-m \tag{25}$$

Clearly, if $V_g-nV^*\cos\left(\bar{\delta}-\delta_g\right)\geq 0$, the system will be stable in the grid-connected mode, and that the stability condition is independent of the transmission line impedance and load.

## III. Simulation Results

TABLE I
PARAMETERS FOR SIMULATIONS

| Parameter | Values | Parameter | Values |
|---|---|---|---|
| $V_g$ (V) | 315 | $Z_{line}$ ($\Omega$) | j0.314 |
| $f^*/f_i$ (Hz) | 50/[49, 51] | $V^*$ (V) | 315/4 |
| $m$ | 0.5 | $\varphi^*$ | 0.2 |

To verify the effectiveness of the power factor angle droop control, simulations are performed on Matlab/Simulink platform. The related parameters of the tested system comprised of four DGs are listed in Table I.

### A. Case 1: Unified control

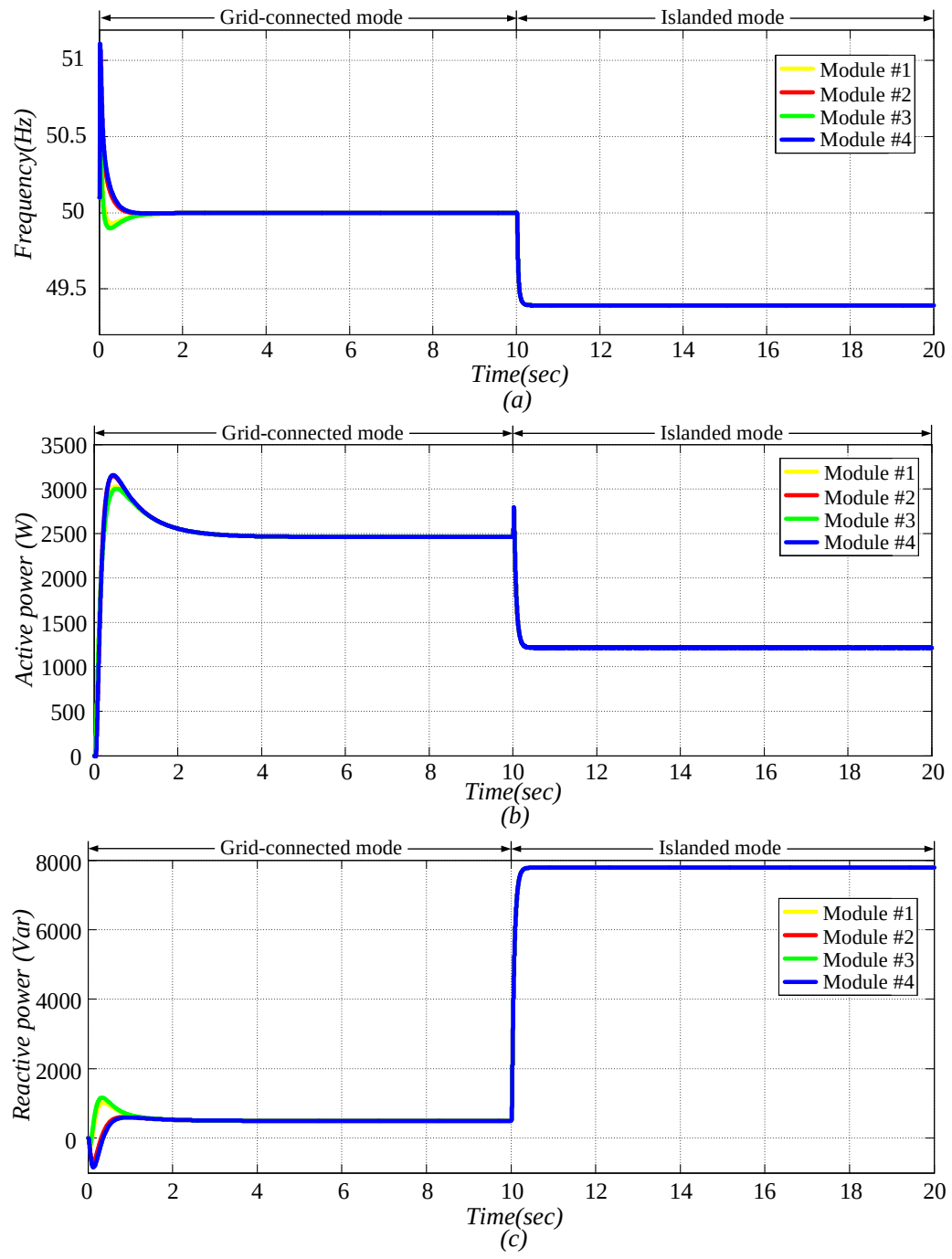

Fig. 2. Simulation results of case 1.

This case is carried out while switching from grid-connected mode to islanded mode. The frequencies of inverters are shown in the Fig. 2(a), which indicates that the proposed scheme can realize the seamless transition between the two modes. The active and reactive power allocations are shown in Fig. 2(b) and (c). Therefore, the proposed scheme is a unified control approach for cascaded inverters to work in the grid-connected and islanded modes.

### B. Case 2: Suited for all types of loads

In this case, the operation in islanded mode is tested under pure resistance, resistance-inductance and resistance-capacitance loads. During interval [0s, 6s], a resistance load is connected to the cascaded inverters; In interval [6s, 12s] a resistance-inductance load is switched into it; After that, the load is changed into a resistance-capacitance load. Fig. 3 shows the waveforms of frequency, active and reactive power of all modules from top to bottom. As seen, the frequencies of all the modules converge quickly after start up, then they are always synchronous no matter how the loads change. However, the frequency changes with load power factor. The frequency under resistance-capacitance load is higher than that under resistance-inductance load, which is in agreement with what equation (7) implies. Meanwhile, the performance in active and reactive power sharing is excellent. As seen, the proposed scheme is suited for all types of loads.

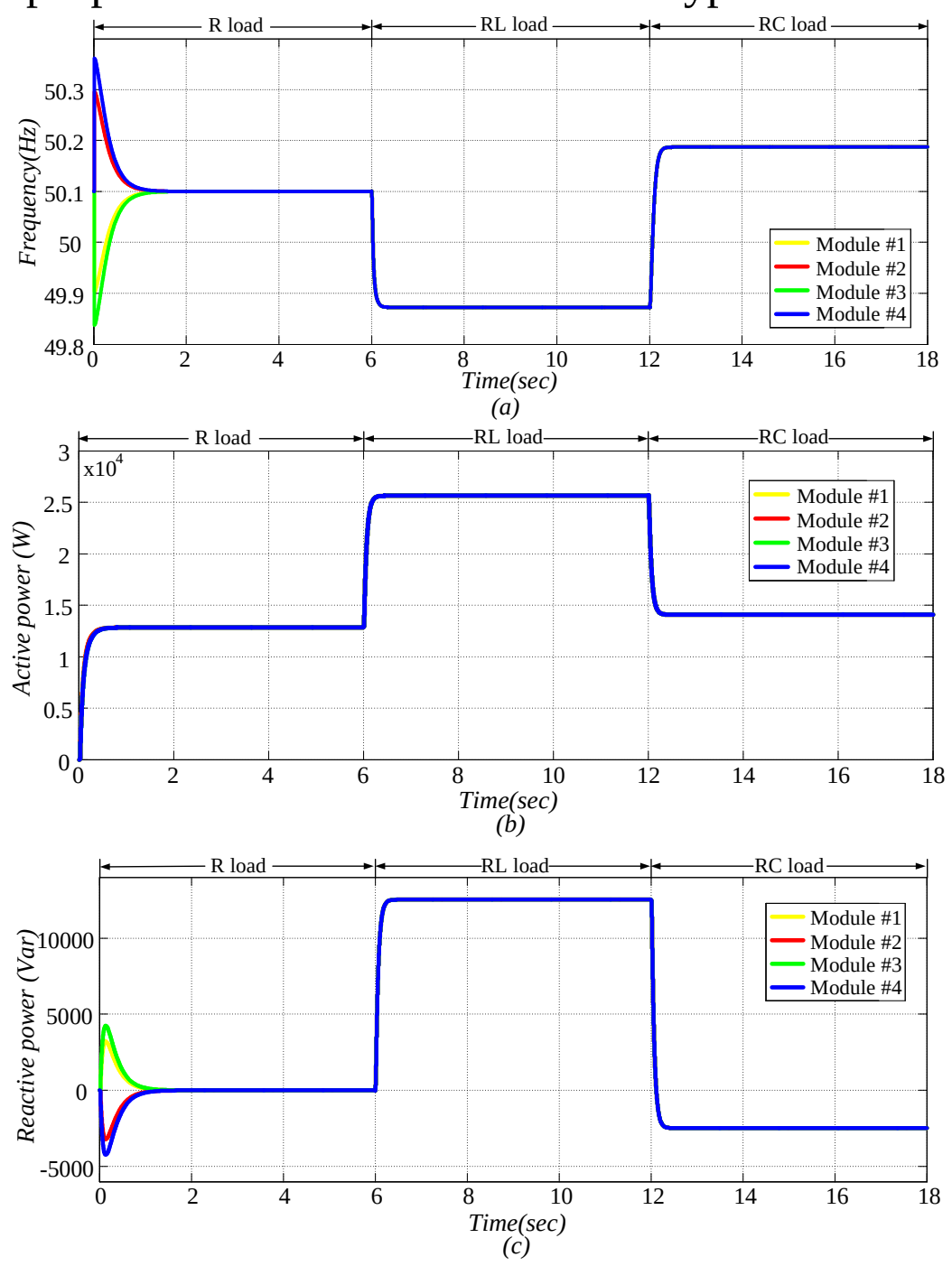

Fig. 3. Simulation results of case 2.

### C. Case 3: Unique equilibrium point

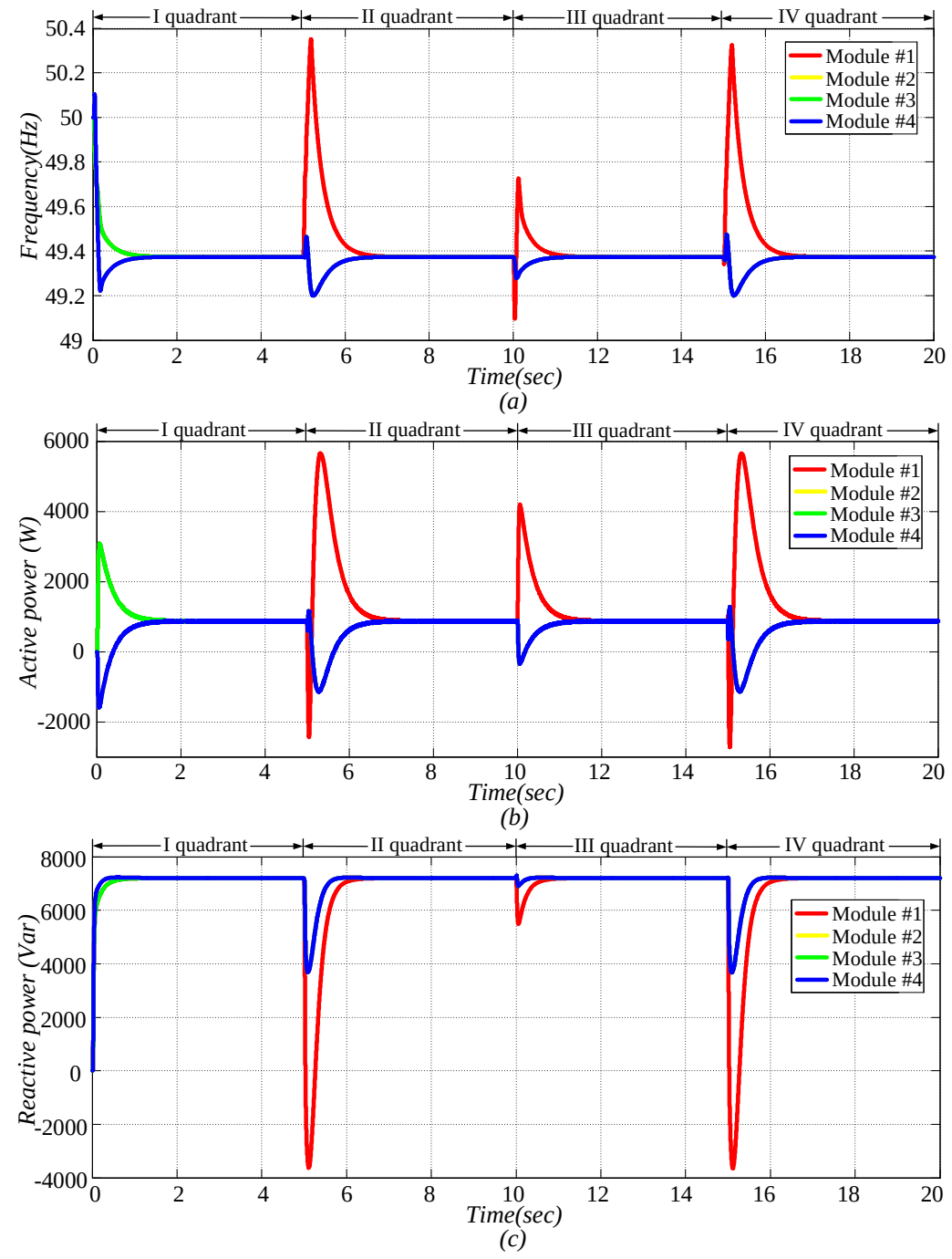

Fig. 4. Simulation results of case 3.

In case 3, the cascaded inverters operate in the islanded mode.

The initial phase angle of inverter #1 is set in I, II, III, IV quadrant in the interval [0s, 5s], [5s, 10s], [10s, 15s] and [15s, 20s], and the initial phase angles of the rest inverters are set as zero. The waveforms of frequency, active power and reactive power are illustrated in Fig. 4(a), (b) and (c), respectively. As seen, the proposed power factor angle droop control scheme always has a unique equilibrium point regardless of the initial states.

### *D. Case 4: Suited for all types of transmission line impedance*

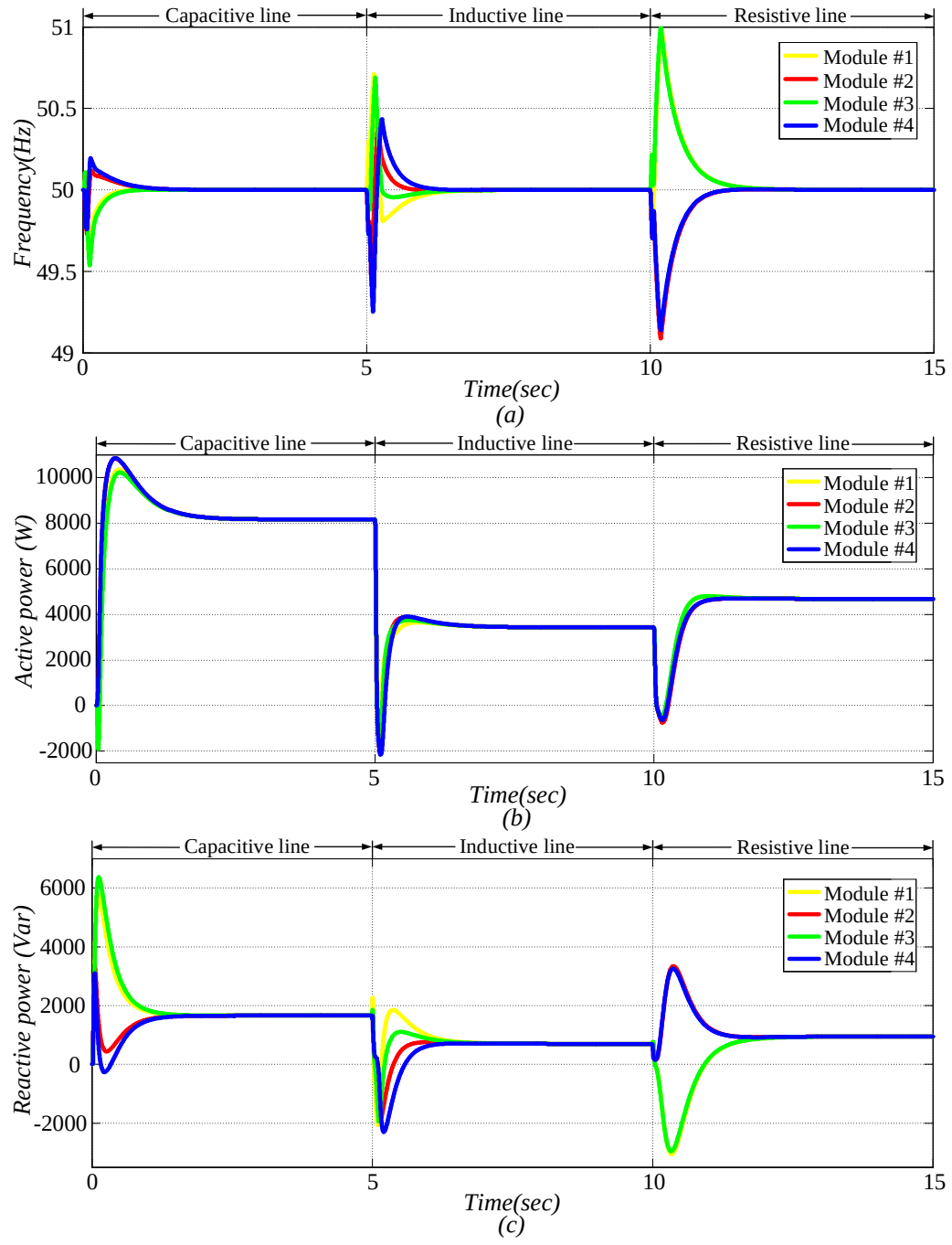

Fig. 5. Simulation results of case 4.

In this case, grid-connected operation is carried out when the capacitive, inductive and resistive transmission line is fed in the interval [0s, 5s], [5s, 10s] and [10s, 15s], respectively. The simulation results of frequencies, active power and reactive power are depicted in Fig. 5 (a), (b) and (c). From the simulation results, the proposed scheme is suitable for all types of transmission line impedance.

### *E. Case 5: Four quadrant operation*

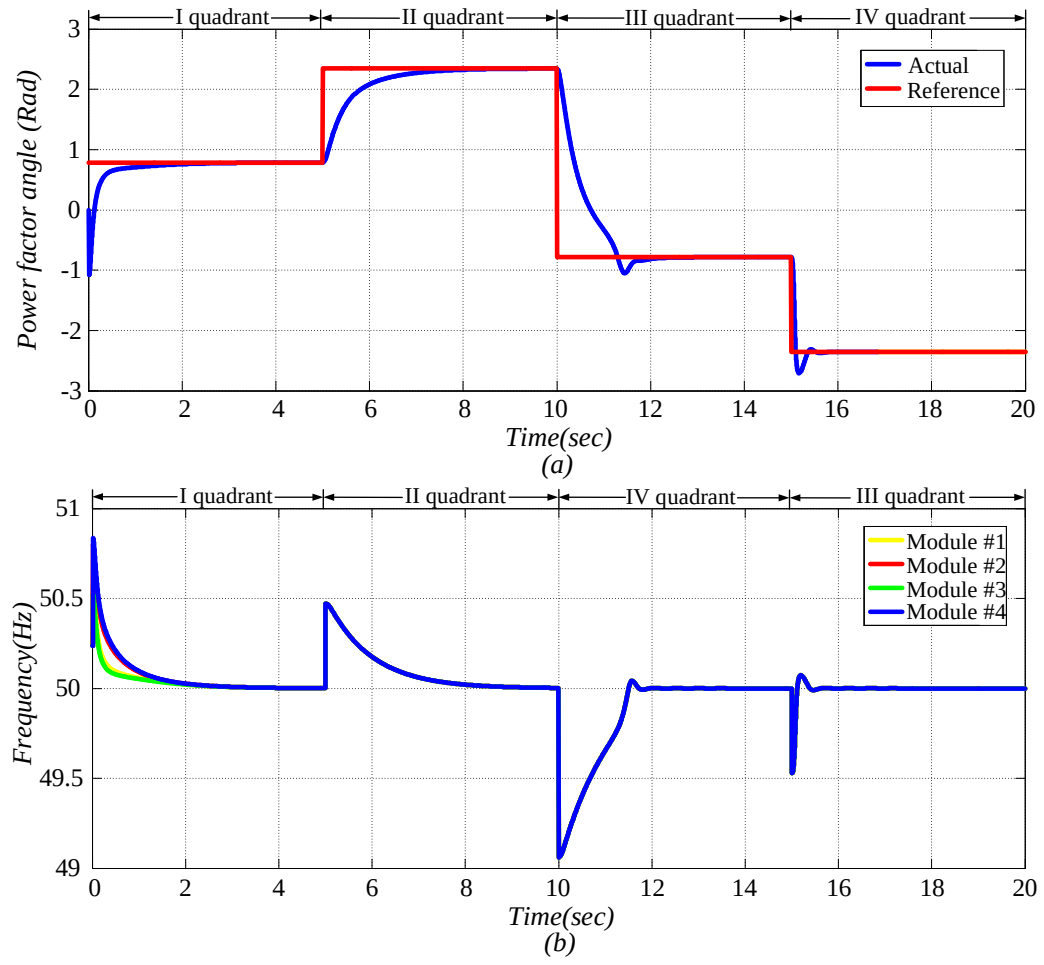

Fig. 6. Simulation results of case 5.

To verify the four quadrant operation capability of the proposed method, grid-connected operation with the power factor angle reference being set as $\pi/4$, $3\pi/4$, $-3\pi/4$ and $-\pi/4$ in the interval [0s, 5s], [5s, 10s], [10s, 15s], and [15s, 20s], respectively, is tested. The waveform of power factor angle is presented in Fig. 6(a), in which the actual one can track its reference. The frequencies are shown in Fig. 6(b), which always converges to 50Hz. Therefore, the proposed scheme can realize the four-quadrant operation.

## IV. Conclusion

This letter proposes a general decentralized control strategy of the cascaded inverters, *i.e.*, power factor angle droop control. It has the following pros: 1) Unified control for both grid-connected and islanded modes; 2) Suitable for all types of loads; 3) Unique equilibrium point; 4) Feasible for all types of transmission line impedance; 5) Four-quadrant operation. Based on the idea behind this study, the decentralized operation of cascaded converters for applications such as PV, storage and STATCOM will be realized.